**Dynamic MR imaging of the skeletal muscle in young and senior volunteers during minimal synchronized neuromuscular electrical stimulation.**

Xeni Deligianni[a,b], Christopher Klenk[c], Nicolas Place[d], Meritxell Garcia[e], Michele Pansini[f], Anna Hirschmann[g], Arno Schmidt-Trucksäss[c], Oliver Bieri[a,b], Francesco Santini[a,b]

[a]*Department of Radiology, Division of Radiological Physics, University Hospital Basel, Petersgraben 4, Basel, Switzerland*

[b]*Department of Biomedical Engineering, University of Basel, Gewerbestrasse 14, Allschwil, Switzerland*

[c]*Department of Sport, Exercise and Health, Division Sports and Exercise Medicine, University of Basel, Birsstrasse 320, Basel, Switzerland*

[d]*Institute of Sport Sciences, University of Lausanne, Bâtiment Synathlon, Quartier UNIL Centre, 1015, Lausanne, Switzerland*

[e]*TMC – European Telemedicine Clinic – a Unilabs company, Torre Mapfre, C/Marina 16 – 18, 08005 Barcelona, Spain*

[f]*Ricerche Diagnostiche Srl, Largo Ignazio Ciaia, 13, Bari, Italy*

[g]*Department of Radiology, University Hospital Basel, Petersgraben 4, Basel, Switzerland*

Original Submission to *Magnetic Resonance Materials in Physics, Biology and Medicine* before Peer Review

**Corresponding author:**

Xeni Deligianni
Division of Radiological Physics, Department of Radiology, University Hospital Basel,
Petersgraben 4, 4031, Basel, Switzerland
E-mail: xeni.deligianni@unibas.ch, Phone:   +41-61-556-5728

*ACKNOWLEDGMENTS*

***This is a pre-print of an article published in Magnetic Resonance Materials in Physics, Biology and Medicine. The final authenticated version is available online at: https://doi.org/10.1007/s10334-019-00787-7.***

This work was supported by the Swiss Foundation for Research on Muscle Diseases (SSEM-FSRMM) and Swiss National Science Foundation (grant Nr. 172876).

ABSTRACT

*Object*

Neuromuscular electrical stimulation (NMES)-induced isometric contraction is feasible during MRI and can be combined with acquisition of volumetric dynamic MR data, in a synchronous and controlled way. Since NMES is a potent resource for rehabilitation, MRI synchronized with NMES presents a valuable validation tool. Our aim was to show how minimal NMES-induced muscle contraction characterization, as evaluated through phase contrast MRI, differs between senior and young volunteers.

*Materials and Methods*

Simultaneous NMES of the quadriceps muscle and phase contrast imaging were applied at 3T to 11 senior (75±3 years) and 6 young volunteers (29±7 years). A current sufficient to induce muscle twitch without knee extension was applied to both groups.

*Results*

Strain vectors were extracted from the velocity fields and strain datasets were compared with non-parametric tests and descriptive statistics. Strain values were noticeably different between both groups at both current intensities and significant differences were observed in the regions of interest between the two electrodes.

*Discussion*

In conclusion, NMES-synchronized MRI could be successfully applied in senior volunteers with strain results clearly different from the younger volunteers. Also, differences within the senior group were detected both in the magnitude of strain and in the position of maximum strain pixels.

*Keywords*

phase contrast sequence; age; strain; quadriceps muscle; electrical muscle stimulation

*INTRODUCTION*

The combination of MRI and neuromuscular electrical stimulation (NMES) not only gives information about the magnitude of the muscle response, but also localized feedback about how each part of the muscle reacts [1, 2]. NMES involves the application of a series of intermittent stimuli to superficial skeletal muscles to trigger visible muscle contractions due to the activation of the intramuscular nerve branches [3]. It can be controlled by adjusting the waveform, frequency and amplitude of stimulation [3] and induces a synchronous activation of the motor units. In combination with imaging, this enables the direct assessment of muscle kinematics through MRI, by synchronizing the MR-data acquisition with NMES as recently presented [4]. This method offers three-dimensional data and direct insights into the muscle contraction capabilities in a completely non-invasive way. It is a low-cost and easily-applicable solution and results significantly depend on the applied stimulation current [4]. Although NMES-responses have been investigated with MRI in the past [1, 2, 5], an extensive evaluation of what to expect as a baseline in the MR-based parameters (e.g. normal velocity, displacement maps, strain, strain rate, etc.) and how these parameters can change because of physiological and pathological processes is still missing. Other existing approaches offer data acquisition before and after the scan [2] or a synchronization process that comes from the sequence and not the stimulator [6]. In addition, these methods only focus on $T_2$ mapping [2, 5] or $^{31}P$ spectra acquisition [5, 6]. Yet, the contraction response of a muscle is important and complimentary to characterize its mechanical/elastic capacities.

Comparably to existing approaches for voluntary exercise protocols [7, 8], the suggested method uses velocity information acquired with phase contrast MRI (PC MRI) and provides dynamic muscle images in a similar way to cardiac imaging [9]. Further quantitative evaluation of PC images yields strain maps [7, 10, 11]. While voluntary contraction follows the Henneman size principle (i.e., small motor units are recruited at lower force levels as compared to larger motor units)[12, 13], standard NMES induces a non-selective and mostly superficial random motor unit recruitment, allowing type II muscle fiber recruitment even at low force levels [3, 12, 14–16].

Age-related changes of the skeletal muscle tissue are associated with a reduction of muscle mass (sarcopenia), which is closely associated with a reduced number of motor units [17–19]. It is well accepted that type II (fast-twitching) fibers are the most affected ones [18, 20, 21], which results into muscle fiber grouping, i.e. the reorganization of the remaining fibers in larger motor units [17, 21]. For these reasons, employing NMES as a tool to study muscle fiber alterations in aged muscle can be particularly interesting.

The aim of this study was to investigate velocity imaging in the quadriceps muscle through NMES-synchronized MRI and evaluate the potential differences between senior and young volunteers. While physiological differences are expected between these two population groups [7], the suitability of the proposed stimulation and imaging protocols to highlight such differences cannot be presumed and is thus the primary endpoint of this study.

*MATERIAL AND METHODS*

The study was approved by the local ethics committee and written informed consent was obtained from all individual participants included in the study. Volunteers with a history of heart or kidney disease, cancer and muscle pathology or any operation on the examined lower extremity within the past five years were excluded. A total of 11 healthy senior (mean age: 74.9 ± 3.4 years, range 70-82; mean height: 170 ± 8 cm, 6 male, 5 female) and 6 healthy young volunteers (age: 29.0 ± 6.5 years, (21-35), height: 179 ± 8 cm, 6 male) were included.

*Experimental setting*

An InTENSity Twin Stim III TENS and NMES Combo (Current Solutions LLC, Austin, TX) was used for the stimulation of the quadriceps muscle and 5.1 x 8.9 cm$^2$ rectangular self-adhesive gel-based NMES electrodes (TENSUnits) were attached to the muscle belly as described in [4]. The electrodes were placed at 15 cm distance from each other and 12 cm from the center of the knee joint on the vastus lateralis (VL). The young volunteers were scanned once with the stimulation level set to 18 mA, which was sufficient to achieve muscle twitching without knee extension [4]. For the senior volunteers, the stimulation was applied first at a minimum level to induce a visible contraction within their comfort levels and with the maximum limit set to 22 mA. Five minutes after the maximum applied level, an additional acquisition was obtained at 18 mA for comparison.

A monopolar square wave with frequency set to 150 pulses/s and pulse duration set to 0.3 ms was used for stimulation. The plateau of each contraction lasted 1 s (i.e., 1 s ramp time, 1 s plateau, 1 s ramp down, 2 s relaxation). A second waveform, generated at the beginning of every stimulation cycle, was used for triggering of the MRI acquisition.

*MR acquisition*

The acquisitions were performed on a 3T clinical MRI scanner (MAGNETOM Prisma, Siemens Healthcare, Erlangen, Germany) in the same way and with the same hardware setup as previously described [4]. For the experimental setup, the NMES device was used to periodically stimulate the quadriceps muscle and was synchronized with a single-slice three-directional gradient echo phase contrast (PC) MRI acquisition [4]. A three-directional gradient echo PC velocity encoding sequence was applied. MR acquisitions were performed on a parasagittal slice (through VL and vastus intermedius (VI) muscles) with a spatial resolution of 2.3 x 2.3 x 5 mm$^3$ and a temporal resolution of 42 ms. The velocity encoding was 25 cm/s (repetition time (TR)/echo time (TE) = 10.6/7.21 ms, bandwidth/pixel = 400 Hz/Px, flip angle = 10°, field-of-view = 225 x 300 mm$^2$, 1 k-space line per segment, acquisition time 5 min) and 94 temporal phases were acquired. In total, during the whole image acquisition time approximately 60 contractions were induced.

*Data processing*

The velocity images were elaborated offline with Matlab (The Mathworks, Inc., Natick, MA, USA). Strain tensors were extracted from the velocity fields as described in [7, 10, 22] and subsequently diagonalized to obtain the strain eigenvalues $e_1$. As the acquisition was limited to a single slice, only the in-plane strain tensors could be extracted from the velocity field.

The post-processing analysis was performed initially for the VL and VI and then four different regions-of-interest (ROI) were selected equidistantly covering both VL and VI muscles proximally to distally in respect to the knee. ROIs 2 & 3 were located approximately in between the two stimulation electrodes (see Figure 1). For every time frame, the spatial median values were calculated (to account for the skewness of the statistical distribution of the values inside the ROIs). Temporal local maximum values were calculated for the strain over each ROI [4].

In addition to the magnitude of the deformation as described by the strain values, temporal information (i.e., the rate of reaching the maximum response) was also extracted from the datasets. This information was obtained in terms of "increase rate" of the strain following the stimulus and was calculated as the slope of the line connecting the beginning of the contraction (defined as the point of maximum curvature of the strain curve) and the maximum point of the same curve. This parameter was defined as strain increase rate and it was descriptively evaluated through maps.

Statistical analysis

Comparison with a significance level of 0.05 was performed between the independent groups of the results from the senior volunteers (SV) for 18 mA and 22 mA (SV18 and SV22) versus the results of the young volunteers (YV) at 18 mA (YV18). The comparison was performed for all four ROIs. The internal control for distribution normality was performed with qualitative histogram visualization. Given the low number of participants non-parametric statistics were applied; since the distributions were not all normal and the number of samples was small, two-sided Wilcoxon rank sum test was used. Statistical analysis was performed with Matlab (*ranksum* function). Due to the small number of volunteers, no statistical analysis between genders was performed, but the results of the male volunteers were analyzed separately to ensure there was no considerable bias (6 young versus 6 senior participants).

*RESULTS*

All six YV were scanned successfully with the stimulation current set at 18 mA. Dynamic PC images were successfully acquired from SV at 18 – 22 mA. For the majority of SV (9 out of 11), a current amplitude of 22 mA had to be applied to achieve a similar muscle twitch as compared to 18 mA in the YV. Two SV were scanned at 20 mA instead of 22 mA, because the lower current already achieved sufficient muscle twitch. Since these were only two cases, the results from the scan at 20 mA of these two SV were not analyzed, but only the ones at 18 mA.

In general, the velocity averaged over e.g. the VL as a function of time presents two pronounced peaks one at the beginning of the contraction and one at the moment of the release of the muscle [4]. In Figure 2, some exemplary velocity vector maps from the beginning of the contraction are presented. The three-dimensional colored velocity vectors from the VL were overlayed on an anatomical image of the thigh. In the senior volunteers, the contraction peak occasionally appeared at a later time frame than in the younger volunteers (i.e. around the $40^{th}$ frame instead of the $30^{th}$ frame).

The principal strain maps were calculated, and the temporal evolution of strain was analyzed for the VL and VI (see Figure 3). As expected, we observed a response to the stimulation in both muscles, the VL and the VI, yet the response in the VL was stronger. When applying a lower current (i.e., 18 mA) to the senior volunteers, there was no discernible response in the VI. Moreover, for the younger volunteers, the strain reaches a maximum value faster than for senior volunteers (see Figure 3).

The strain values in both VL and VI were summarized for the four different ROIs (ROI1-4: proximal to distal). As expected, for the central ROIs 2 & 3, that are located approximately between the electrodes, the strain values had significantly lower values for the senior in comparison to the younger volunteers (Figure 4).

Finally, the analysis of the spatial distribution of the strain increase rate showed various different patterns. Figure 5 shows three out of five cases of senior volunteers with no distinct regions characterized by higher values (Figure 5d, 5f, 5h) and three out of four cases, in which those regions were very small (Figure 5g, 5i, 5j). On the contrary, some distinct connected regions with higher rates were found for all young volunteers and four of the senior volunteers (Figure 5a, 5b, 5c, 5e).

*Statistical analysis*

The results of the two-sided Wilcoxon rank sum test are presented in Table 1. The comparison was performed for the four ROIs of both VL and VI. The differences between young and senior volunteers for contractions at 18 mA were significant for the 2 central ROIs ($p$=0.002). Significant differences of strain values from young volunteers and senior volunteers at 22 mA were found for ROI 3 ($p$=0.012), which is in agreement with the boxplot visualization (see Figure 4).

Comparison of male volunteers (6 young and 6 senior) revealed different range and *p*-values (see Table 1), however, the result of the test remained the same.

*DISCUSSION*

The aim of this study was to investigate velocity imaging in the quadriceps muscle through NMES-synchronized MRI and evaluate the potential differences between senior and young volunteers. Significant differences in skeletal muscle contraction parameters (i.e., principal strain) were assessed with PC MRI between healthy young (<35 years old) and senior (>70 years old) volunteers. It was also shown that the difference of strain values between the young and senior volunteers, at the same stimulation current, was larger at the ROIs between the two NMES electrodes.

The significant difference in strain values between the two age groups is in agreement with Sinha et al. who showed differences not in strain but in strain rate maps calculated from PC images in senior and younger volunteers (78 years vs 28 years) during voluntary contractions [7]. While this agreement seems straightforward, the results presented in this work could not be simply deduced from similar data acquired during voluntary contraction, since the two types of exercise are fundamentally different. In our case, the difference between the two populations could be attributed to stiffer muscle elasticity with increasing age [17, 18, 23]. A reduced number and size of mainly type II muscle fibers in the seniors might contribute to a difference in strain values between young and senior individuals as well, since NMES allows type II muscle fiber recruitment even at low force levels [17, 18, 20].

In addition, the spatial distribution of the parameter of the strain curve, defined here as "strain increase rate", was examined. This parameter intimately relates to the strain rate, determined as the temporal derivative of strain, but it still depends on the reference state. For linearly increasing strain curves, the two parameters should be approximately alike. However, the muscle response to electrical stimulation is not linear (i.e., the force, the magnitude of stretching, etc) and often not monotonic and thus this assumption is typically not valid. This was the case especially in the senior volunteers, who overall proved to be less responsive to the same stimulation current.

In the present study, we observed a faster response of higher amplitude, that "activated" a larger area of the most superficial muscles in younger compared to the senior volunteers. This observation can be used as a potential marker to show efficacy and improvement of NMES training protocols in the aged muscle. Furthermore, within the SV group some responses were similar to the ones of the YV group. A next step would be to investigate whether this fact depends on special characteristics of one's physical status, which has to be characterized with other parameters such as external force measurements.

Clinical applications of the presented method include a variety of pathological muscle conditions sensitive to fiber type II atrophy such as chronic obstructive pulmonary disease [24], or chronic steroid myopathy [3, 25], and age-related diseases such as sarcopenia [26]. It can also be applied to competitive or elderly people who need to train fast fibers with low effort [15, 16].

Due to the limited number of volunteers in this study, the group of senior volunteers was considered as one single group. However, there was a variability in the physical condition of the subjects, since some of the senior volunteers performed vigorous training on a regular basis. For better differentiation, the volunteers would have to be grouped more strictly according to their physical condition and training habits, which can be the subject of future investigation.

Finally, one technical restriction of the current study is that there was no comparison of the calculated strain with the force output. In part, this was due to the lack of a suitable measurement equipment at our institution. Yet the choice of using a stimulation current that only generates a visible twitch of the muscle without noticeable knee extension was dictated by the strong discomfort associated with NMES at higher force outputs [3], which makes the detectability of physiological differences at minimal stimulation intensity very relevant for the compliance and comfort of a potential patient. Nevertheless, it would be interesting to compare strain for the same force output in future investigations.

In conclusion, strain measurements with MRI of NMES-induced muscle contraction show age-related differences between healthy volunteers above 70 and below 35 years old. The differences were more significant when the same stimulation current was used for young and senior subjects. Moreover, there were prominent differences not only in the strain magnitude, but also in the temporal rate of strain and variable for different muscle regions. Despite these physiological inter-individual differences, the data shown here may be used as a preliminary data baseline for a more accurate and detailed assessment of muscle function disturbances.

*ACKNOWLEDGMENTS*

This work was supported by the Swiss Foundation for Research on Muscle Diseases (SSEM-FSRMM) and Swiss National Science Foundation (grant Nr. 172876).

*AUTHORS' CONTRIBUTION*

XD: Study conception and design, Acquisition of data, Analysis and interpretation of data, Drafting of manuscript, Critical revision

CK: Study conception and design, Drafting of manuscript

NP: Analysis and interpretation of data, Drafting of manuscript, Critical revision

MG: Study conception and design, Acquisition of data, Drafting of manuscript

MP: Study conception and design, Drafting of manuscript

AH: Study conception and design, Drafting of manuscript, Acquisition of data

AS-T: Study conception and design, Drafting of manuscript

OB: Study conception and design, Drafting of manuscript

FS: Study conception and design, Acquisition of data, Analysis and interpretation of data, Drafting of manuscript, Critical revision

*Conflict of Interest*:

The authors declare that they have no conflict of interest.

*Ethics:*

The study was approved by the local ethics committee and written informed consent was obtained from all individual participants included in the study

TABLES

|  | | YV18-SV18 | | | YV18-SV22 | | |
| --- | --- | --- | --- | --- | --- | --- | --- |
|  | Gender | R | p-value (CI95%) | H | R | p-value (CI95%) | H |
| ROI-1 | m/f | 71 | 0.089 | 0 | 52.5 | 0.634 | 0 |
| ROI-2 | m/f | 83.5 | 0.002 | 1 | 59 | 0.212 | 0 |
| ROI-3 | m/f | 83.5 | 0.002 | 1 | 68.5 | 0.012 | 1 |
| ROI-4 | m/f | 69 | 0.142 | 0 | 56 | 0.372 | 0 |
| ROI-1 | m | 51 | 0.065 | 0 | 38.5 | 0.701 | 0 |
| ROI-2 | m | 55 | 0.011 | 1 | 35 | 0.892 | 0 |
| ROI-3 | m | 57 | 0.002 | 1 | 47.5 | 0.048 | 1 |
| ROI-4 | m | 49 | 0.128 | 0 | 40.5 | 0.455 | 0 |

Table 1. Results of Wilcoxon rank sum test for strain between young volunteers' values at 18 mA (YV18) and senior volunteers at 18 and 22 mA (SV18 and SV22) (*R*: order of significance, *p*: *p*-value of the test, H=0: the null hypothesis cannot be rejected at the 5%, H=1 indicates that the null hypothesis can be rejected at the 5% level).

FIGURE LEGENDS

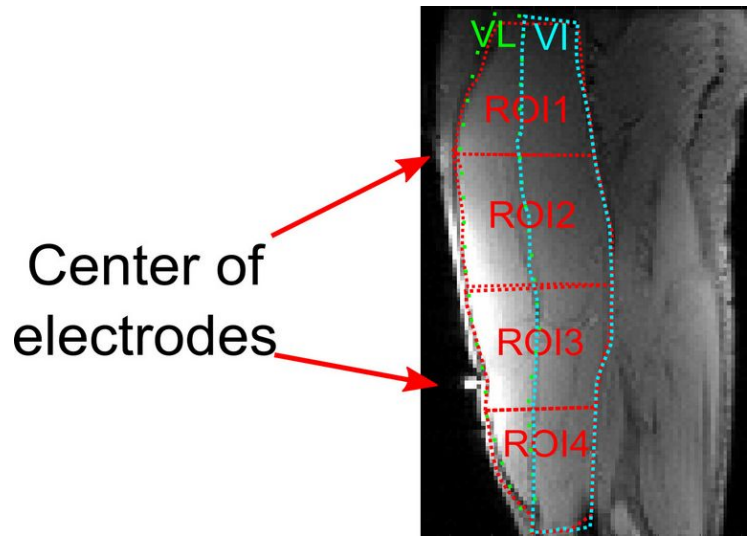

Fig. 1 Description of the placement of the regions of interest in respect to the muscles and to the electrodes position

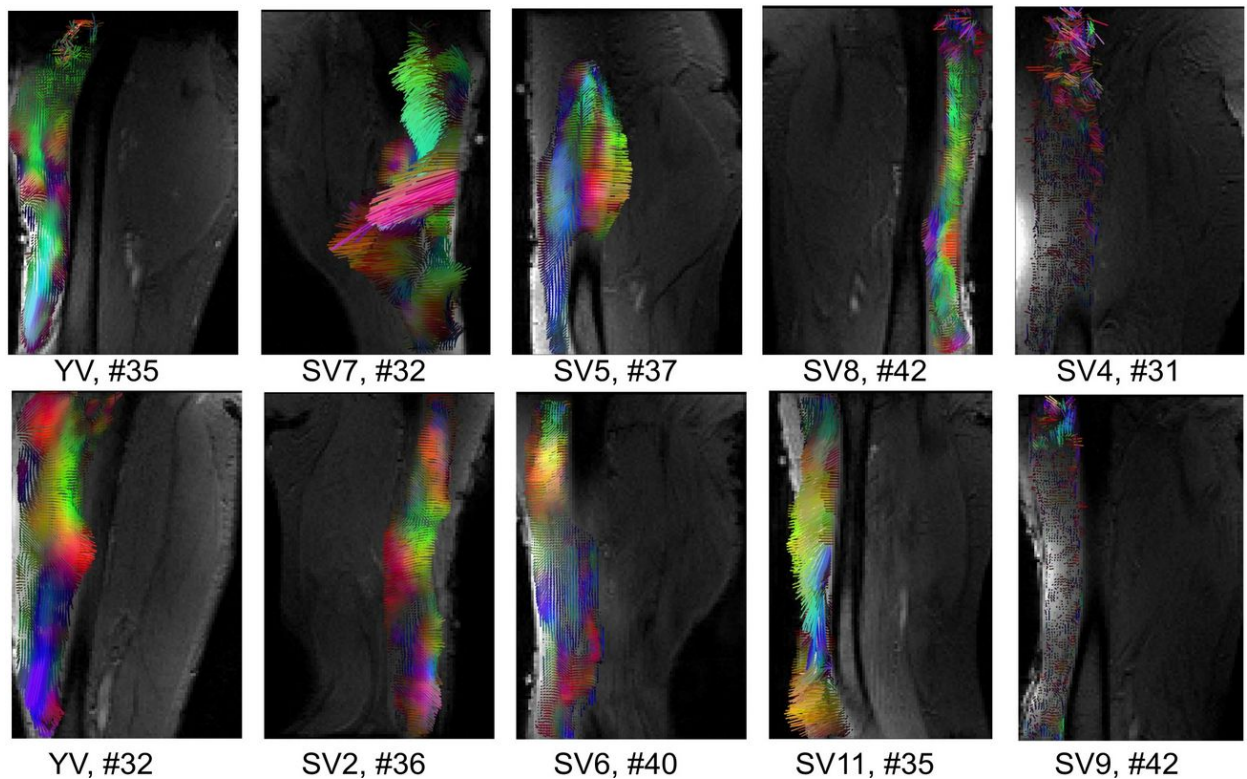
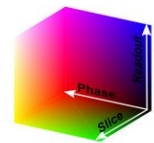

Fig. 2 Three-dimensional velocity maps at the beginning of the contraction plateau for ten exemplary cases of young and senior volunteers. The colored vectors are normalized according to the maximum. Velocity vectors inside the quadriceps are represented as color-coded lines (i.e., as in the cube) according to the direction (*red*: phase encoding direction, *blue*: readout direction, *green*: slice direction). The difference in homogeneity of the contraction can be seen in the color-coding

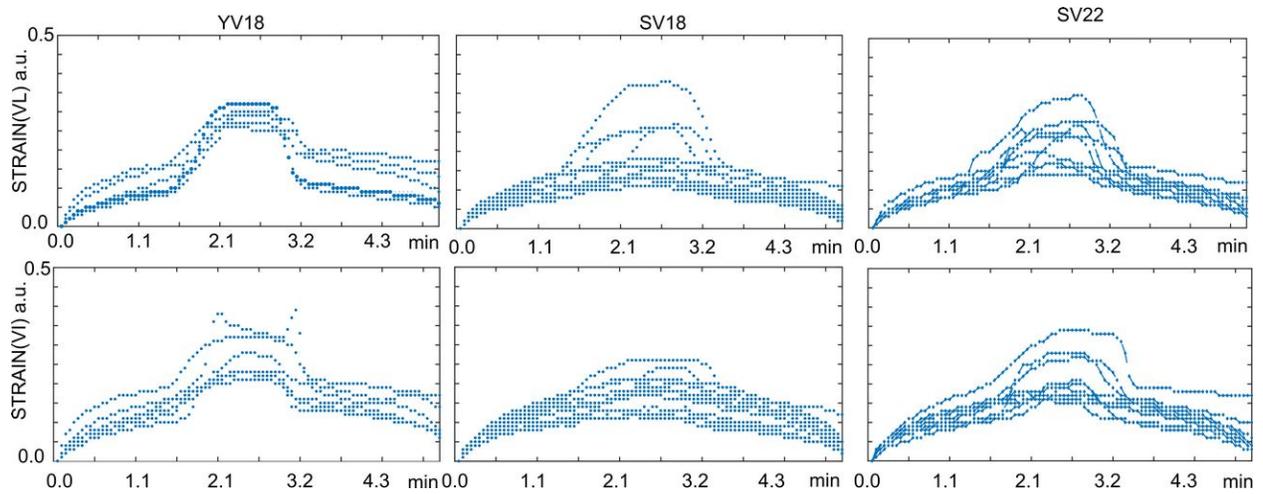

Fig. 3 Temporal evolution of strain calculated for the vastus lateralis (VL, *upper row*) and vastus intermedius (VI, *lower row*) at every time frame given in arbitrary units (*a.u.*). Results are given for both young (YV18: young volunteers at 18 mA) and senior volunteers (SV18: senior volunteers at 18 mA, SV22: senior volunteers at 22 mA). Large variability both in maximum values and breadth of the curve can be observed

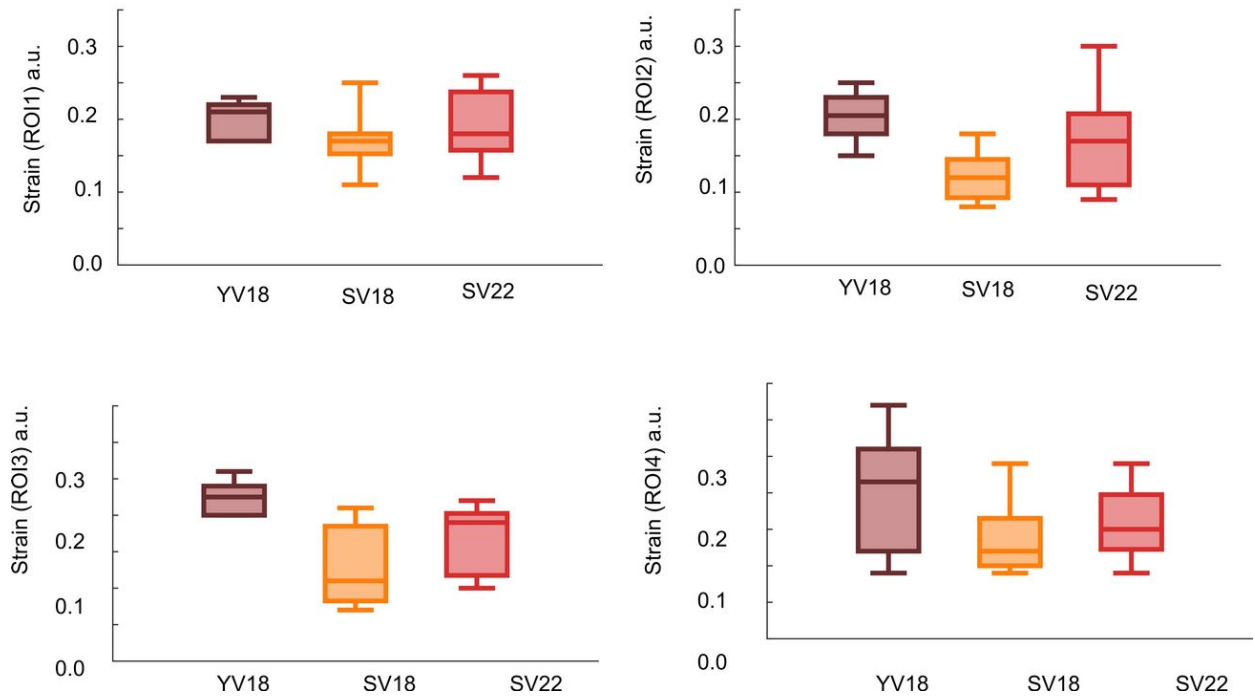

Fig. 4 Boxplots of strain values of the young volunteers (YV) at 18 mA (YV18, *left boxplot*), and of the senior volunteers (SV) at 18 mA and 22 mA (SV18, *central boxplot* and SV22, *right boxplot*) averaged over four different ROIs of the vastus lateralis from proximal to distal ($ROI_1$ to $ROI_4$ see Fig. 1)

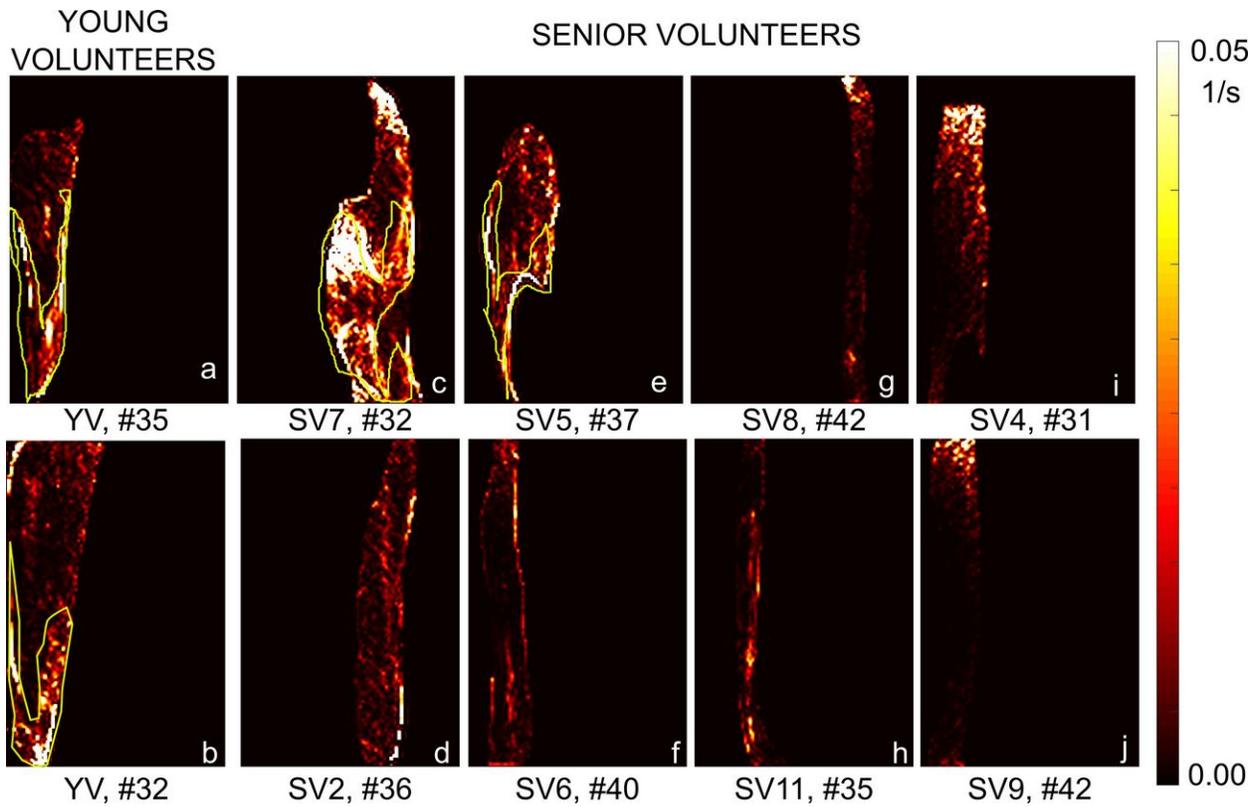

Fig. 5. Examples of strain increase rate maps of young volunteers (YV-*left*) and senior volunteers (SV-*right*). The respective number of the time frame is given on every image (YV/SV number, # *number of frame*). A clear region with hyperintense values at the borders of the muscle (see delineation in green) can be identified in both young volunteers, but only in a limited number of two senior subjects